\documentclass{aastex61}

\newcommand\aastex{AAS\TeX}

\shorttitle{\aastex\ sample article}

\begin{document}
\title{Discovery of A Compact X-ray Object with A 614\,s Periodicity in the Direction of the Galactic center}

\correspondingauthor{Hang Gong}
\email{ghang.naoc@gmail.com}

\author{Hang Gong}
\affil{CAS Key Laboratory of Optical Astronomy, National Astronomical Observatories, Chinese Academy of Sciences, Beijing 100101, China}




\begin{abstract}
We report on analysis of X-ray, optical and radio observations of the previously overlooked X-ray source 2CXO\,J174517.0$-$321356 located just 3.2$^{\circ}$ away from the Galactic center. Timing analysis of X-ray observations of the source with \textit{XMM-Newton} reveals periodic pulsations with periods of 1228\,s and 614\,s, with the latter being tentatively considered fundamental. On the other hand, an observation of the object with \textit{NuSTAR} reveals hard thermal-bremsstrahlung spectrum. Inspection of the archival VLT image reveals, however, no obvious optical counterpart down to $\rm{R}>25\,$mag. Observations made with ATCA showed a possible faint radio counterpart with a positive spectral index ($\alpha > 0.51$) between 1--3\,GHz, but follow-up ATCA and VLA observations at frequencies between 4.5--10\,GHz and 3--22\,GHz, respectively, could not detect it. Given the properties in these three bands, we argue that the most likely origin of the X-ray source is emission from a new intermediate polar close to the Galactic center. Alternatively, and less likely, it is an ultra-compact X-ray binary, which is one of the most compact X-ray binaries.

\end{abstract}


\section{Introduction}
Identification of multiple serendipitously discovered X-ray sources in crowded regions of the Galactic plane, and especially the center is notoriously difficult \citep{2016ApJ...825..132H}. This is due to the fact that most of the sources are faint in X-rays which precludes sensitive searches for possible pulsed signal and detailed broadband spectral studies. Therefore, X-ray data alone is most often not sufficient to establish the origin of the source. Additionally, density of optical and infrared sources together with high absorption in the Galactic plane make it complicated to identify and study properties of potential counterpart sources. In fact, even detection of periodic signals in the X-ray band is often insufficient to establish the origin of an X-ray source as characteristic spin and orbital periods of NS (neutron star) and WD (white dwarf) X-ray binaries partly overlap (hundreds of seconds to thousands of seconds).

For example, 4U\,1820$-$30 is the most compact X-ray binary so far, but the orbital nature of 4U\,1820$-$30's 685\,s signal discovered in the \textit{EXOSAT} data was not obvious. It was deduced based on its high $L_{\mathrm{X}}$, period stability within one year and binary evolution models \citep{1987ApJ...312L..17S}. Another example is the most compact binary known to date RX\,J0806.3$+$1527. Its 321\,s signal detected in the \textit{ROSAT} data was mistakenly treated as a rotational period \citep{1999A&A...349L...1I} before its orbital nature was deduced from a series of spectroscopic observations \citep{2002A&A...386L..13I,2010ApJ...711L.138R}.
Misidentifications of rotational periodic signals occurred too. For example, although initially different scenarios \citep{2010A&A...523A..50F,2010MNRAS.408..975M} were suggested, AX J1740.2$-$2903 (or 2XMM J174016.0$-$290337) was finally found to be an intermediate polar (IP) by optical spectroscopic observations \citep{2012A&A...538A.123M,2010ATel.2664....1H}. It is located 1.2$^{\circ}$ away from the Galactic center and has a 626\,s modulation in its \textit{XMM-Newton} light curve. Another example is OGLE-UCXB-01 \citep{2019ApJ...881L..41P}, which is also suspected to be an IP \citep{2021ApJ...916...80P}, despite its name.

Here, we report a newly discovered periodic variability with period 614\,s from a previously overlooked X-ray source 2CXO\,J174517.0$-$321356 \citep[R.A., Dec=$266.32087^\circ,\ -32.23237^\circ$; J1745$-$3213 hereafter,][]{2019HEAD...1711251G,2019HEAD...1711401E} which is 3.2$^{\circ}$ away from the Galactic center. By analysing its X-ray, optical and radio data, we argue the signal is probably due to a rotational period making it one of the IPs with confirmed periods in the direction of the Galactic center, or less likely, an orbital period making it one of the most compact X-ray binaries \footnote{\url{https://personal.sron.nl/~jeanz/ucxblist.pdf}}.

\section{X-ray}
\subsection{Archival X-ray Data Analysis}
J1745$-$3213 was serendipitously covered by a \textit{Chandra} observation (ObsID=13577) and an \textit{XMM-Newton} observation (ObsID=0553950201) before 2021 (Table.1 and the left panel of Fig.\,1). The 2\,ks \textit{Chandra}/ACIS observation collected about 70 counts (0.3--8\,keV, off-axis angle $\approx 3.7'$, $\sigma>27$) within the 3$\sigma$ detection region given by \textit{wavdetect} \citep{2002ApJS..138..185F}. The \textit{XMM-Newton} observation was made for black hole binary H1743$-$322, which is $12'$ away from J1745$-$3213. Due to the large off-axis angle, the EPIC-PN detector, used in small-window mode, did not cover J1745$-$3213. As Fig.1 (\textit{left}) shows, the X-ray source is significant on the EPIC-MOS detectors.

\begin{table*}
\begin{center}
\caption{Archival Observations}
\begin{tabular}{llll}
\hline\hline
Telescope  &Instrument& Date   & Exposure Time      \\
 \hline
\textit{XMM-Newton}&MOS& 2010-10&86\,ks\\
 \hline
\textit{Chandra}&ACIS-I& 2012-02&2.0\,ks \\
\hline
\textit{NICER}&/&2019-09--2019-10&15\,ks\\
\hline
VLT&VIMOS&2013-04& 120\,s\\
\hline
\end{tabular}
\end{center}
\end{table*}

\begin{figure*}
\centering
\includegraphics[width=0.4\textwidth]{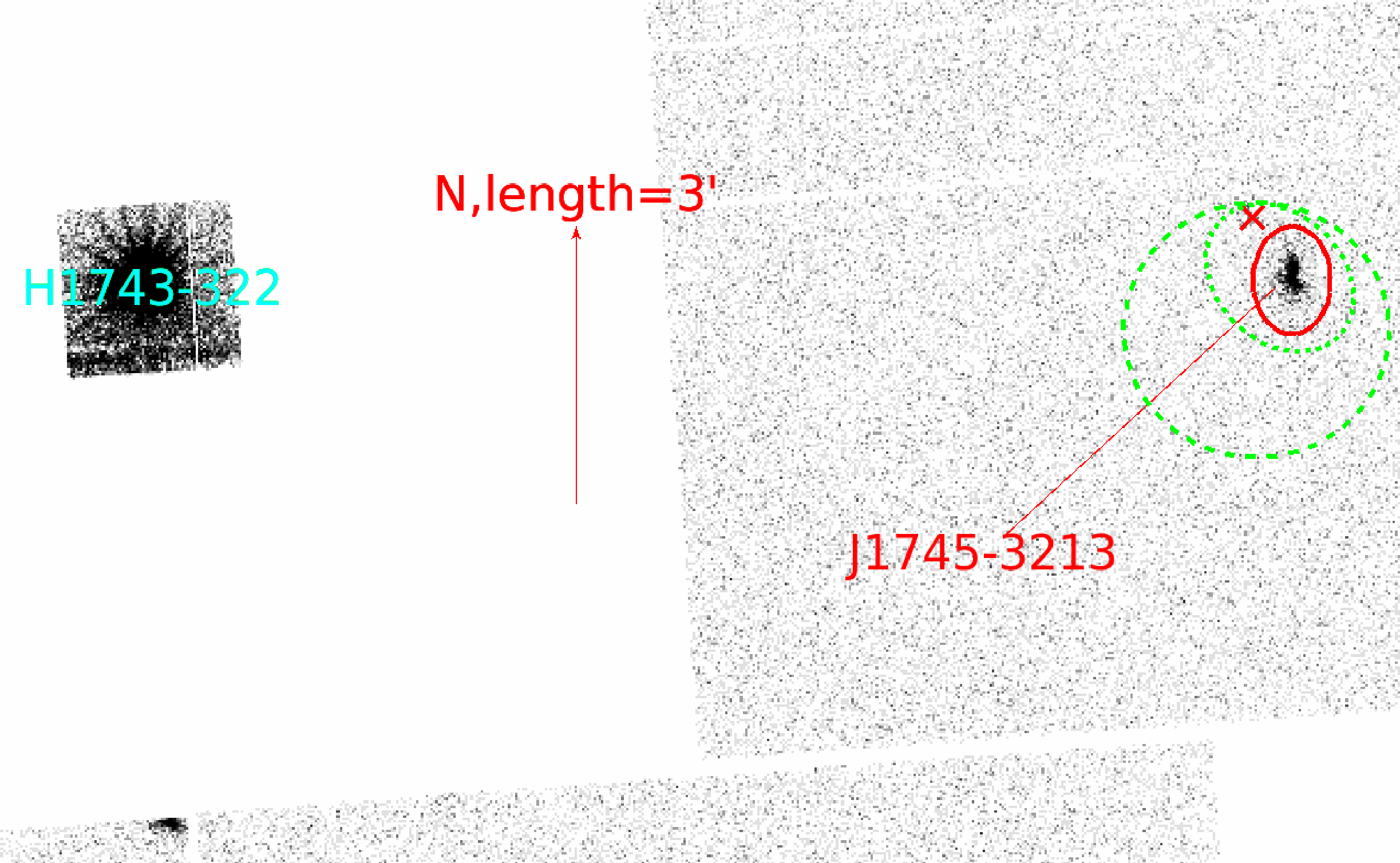}
\includegraphics[width=0.4\textwidth]{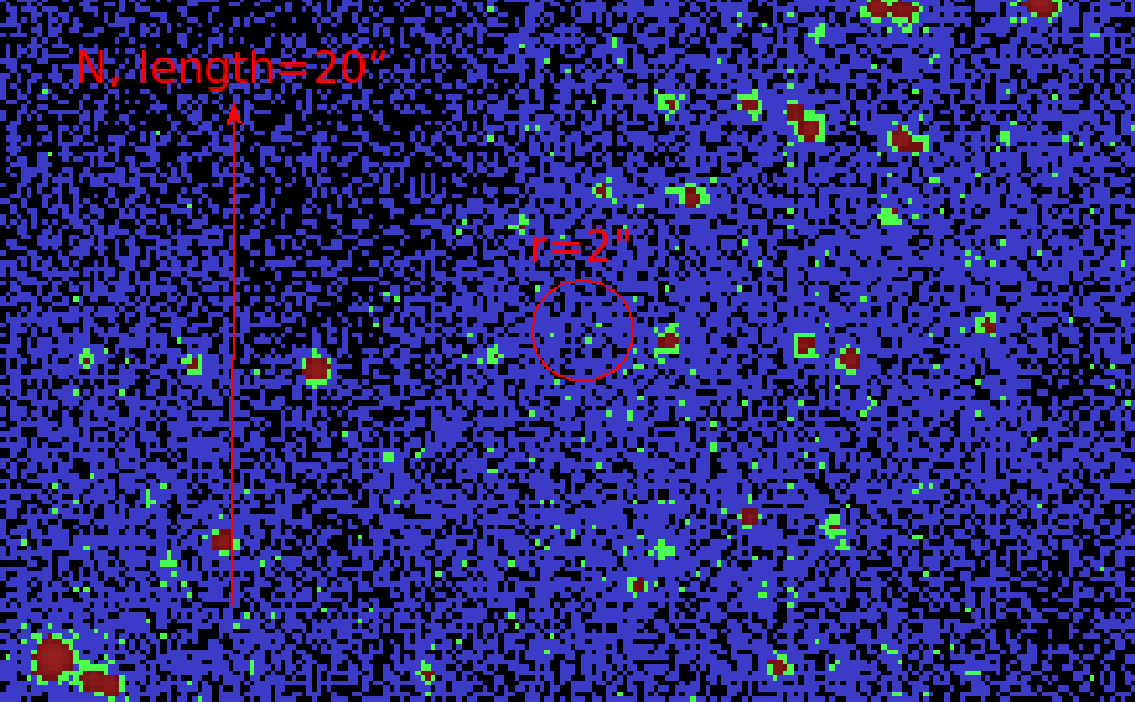}
\caption{Left panel: The red ellipse ($30"$x$40"$) marks the source extraction region between 0.3--10\,keV for EPIC-MOS. We selected the region between the green dashed lines for the background extraction, which also avoids the source marked with an X. Right panel: The VIMOS R-band image of J1745$-$3213. Bias, flat-field and astrometric corrections have been applied. Color scale is fine-tuned in DS9 to highlight faint features. The red circle indicates the \textit{Chandra} position.}
\end{figure*}
\subsubsection{Spectral Analysis}
For \textit{XMM}, the MOS data were reduced by {\sc SAS} 19 \citep{2004ASPC..314..759G} with standard procedures. The two event files of MOS1 and MOS2 were generated by \textit{emchain}, filtered by \textsc{gti} files, which were generated by \textit{espfilt} automatically, and then barycentric corrected by \textit{barycen} and DE405 ephemeris. Source photons were extracted between 0.3--10\,keV using \textit{evselect} around the position given by \textit{Chandra}. Due to the large off-axis angle\footnote{\url{https://xmm-tools.cosmos.esa.int/external/xmm_user_support/documentation/uhb/offaxisxraypsf.html}} of J1745$-$3213, which causes elongated images of point sources on the MOS detectors, and the existence of a source nearby (marked by the red X in the left panel of Fig.\,1), we tried different regions and finally adopted an elliptical region with semi-major and semi-minor axes of $\rm{r_{a}}\times r_{b}$=$30''\times40''$ (MOS1+MOS2$\approx$3,500 photons) as the source extraction region, and a non-concentric ring as the background region (Fig.\,1, \textit{left}).

We used {\sc Sherpa} of {\sc CIAO} 4.14 \citep{2006SPIE.6270E..1VF} to perform the spectral fitting. Every spectral bin had 25 counts at least. Hence, we adopted \textit{chi2datavar} instead of the default \textit{chi2gehrels} as the fit statistic\footnote{\url{https://cxc.cfa.harvard.edu/sherpa/statistics}}. The background was subtracted channel by channel using \textit{subtract}. The two MOS spectra were fit simultaneously. Among the commonly used models\footnote{\url{https://cxc.cfa.harvard.edu/sherpa/models/}}, we found a simple \textit{apec}, \textit{thermal-bremsstrahlung}, \textit{thermal-bremsstrahlung} with a partial covering absorber \citep[e.g.,][]{2002A&A...387..201H} and \textit{diskbb} led to poor or unreasonable fits to the data. Among most of the models used, residuals around 6.7\,keV suggested presence of an iron feature which was thus included in the fits (Fig.\,2, \textit{left}). Due to its broadness ($\approx$200\,eV), the feature is likely to be a mixture of three kinds of iron lines (6.4, 6.7 and 7.0\,keV). Nevertheless, adding three Gaussian components made the fitting complicated and difficult to find a global minimum. We also tried to separate the three possible lines like \citet{2016MNRAS.460..513T} and \citet{2022MNRAS.511.4582C}. However, no matter the three line energies and line widths were fixed or not, our fits could not improve significantly or be reasonable. Thus, we only adopted one single Gaussian component, but left the line width free or fixed it at 0.05\,keV. Hereafter, all the models we mention include an iron feature.

We found the best models (Table.2) include an absorbed power-law (PL) and an inferior absorbed blackbody (BB). If we adopt the absorbed BB as the moderate choice (which is also acceptable in the second \textit{XMM} observation) and its parameters ($N_\mathrm{H}$ and \textit{kT}), the unabsorbed flux and luminosity of J1745$-$3213 between 0.3--10\,keV are about $1.6^{+0.2}_{-0.2}\times\rm{10^{-12}\,erg\,cm^{-2}\,s^{-1}}$ and $1.9^{+0.2}_{-0.2}\times10^{32}\times$(D/1 kpc)$^{2}$ erg s$^{-1}$, respectively.

\begin{figure*}
\centering
\includegraphics[width=0.32\textwidth]{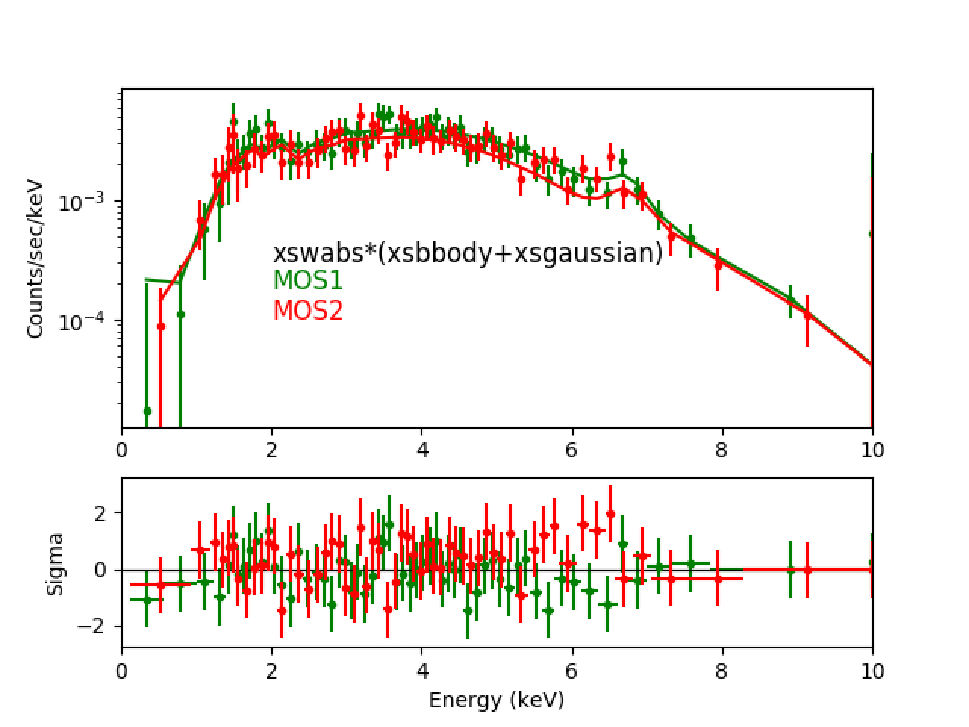}
\includegraphics[width=0.32\textwidth]{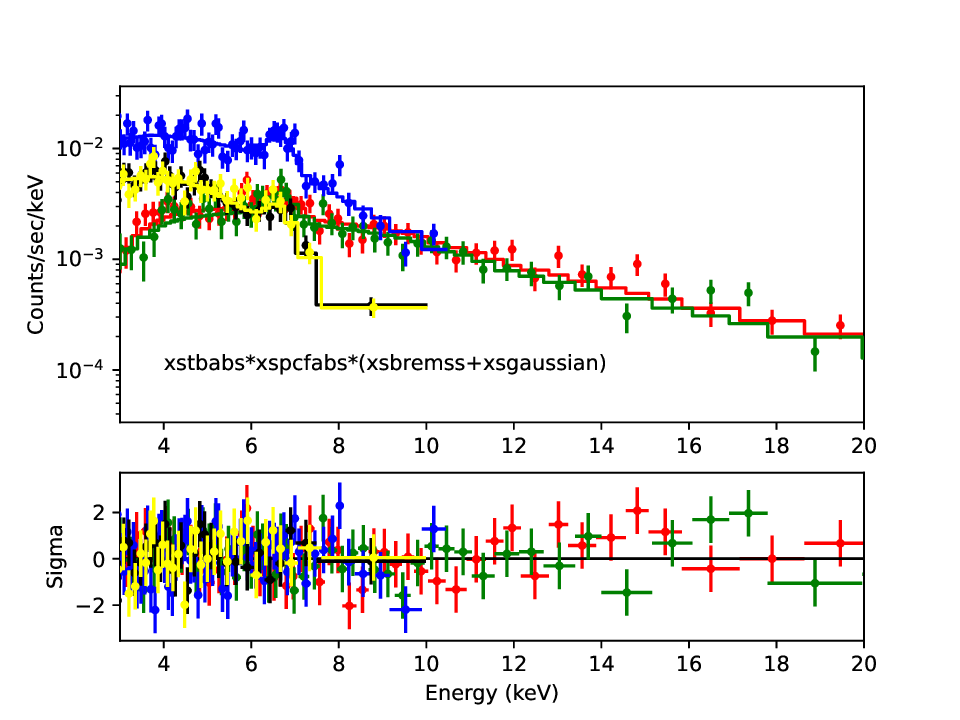}
\includegraphics[width=0.32\textwidth]{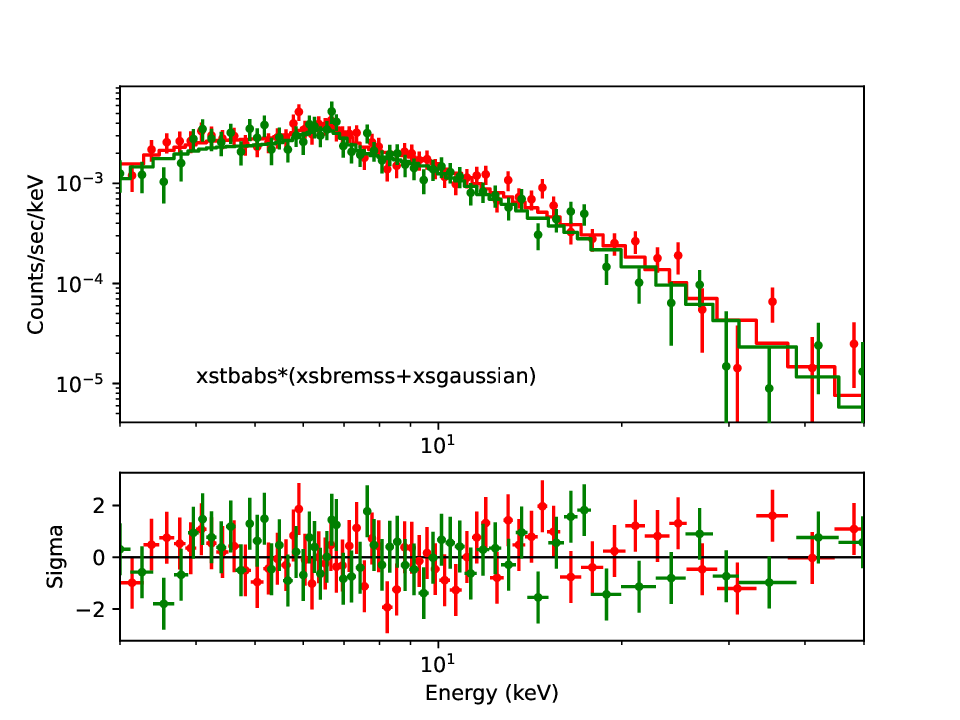}
\caption{Left panel: Two MOS spectra with the BB+Gauss model during the first \textit{XMM} observation. There are obvious residuals around 6.7\,keV. Middle panel: A partial covering \textit{thermal-bremsstrahlung} model is one of the best models that describe the 0.3--20\,keV band during the new observation. The blue, black, yellow, red and green lines represent PN, MOS1, MOS2, FPMA and FPMB, respectively. Right panel: An absorbed \textit{thermal-bremsstrahlung} is the best model to describe the 3--50\,keV band during the new observation. The cross-normalization factor between FPMA and FPMB is $\approx92\%$.}
\end{figure*}

\begin{table*}
\scriptsize{
\caption{X-Ray Spectral Fits Based on Archival \textit{XMM} Data}
\begin{center}
\begin{tabular}{lll}
\hline\hline
Model & Parameters$^{a}$&red.$\chi^{2}$(dof) \\
\hline
xstbabs*(powlaw1d+xsgaussian)&$N_\mathrm{H}$=$1.8^{+0.2}_{-0.2}$,$\Gamma$=$0.5^{+0.1}_{-0.1}$ &1.04(109)\\
&$E_{\mathrm{line}}$=6.7$^{+0.1}_{-0.1}$ ,${\sigma}_\mathrm{line}$=0.2$^{+0.1}_{-0.1}$&\\
\hline
xstbabs*xsbbody&$N_\mathrm{H}$=$0.4^{+0.1}_{-0.1}$,\textit{kT}=$2.7^{+0.1}_{-0.1}$&1.22(112)\\
\hline

xstbabs*(xsbbody+xsgaussian)&$N_\mathrm{H}$=$0.5^{+0.1}_{-0.1}$,\textit{kT}=$2.6^{+0.2}_{-0.2}$&1.18(109)\\
&$E_{\mathrm{line}}$=6.7$^{+0.1}_{-0.1}$,${\sigma}_\mathrm{line}$=0.2$^{+0.1}_{-0.1}$&\\
\hline
xstbabs*xspcfabs*powlaw1d&$N_\mathrm{H}1$=$1.1^{+0.2}_{-0.2}$,$N_\mathrm{H}2$=$8.3^{+2.5}_{-2.5}$&1.22(110)\\
&$f_{\mathrm{CvrFract}}$=$0.6^{+0.1}_{-0.1}$,$\Gamma$=$0.8^{+0.2}_{-0.2}$&\\
\hline
xstbabs*xspcfabs*(powlaw1d+xsgaussian)&$N_\mathrm{H}1$=$1.1^{+0.2}_{-0.2}$,$N_\mathrm{H}2$=$8.1^{+2.2}_{-2.2}$&1.18(107)\\
&$f_{\mathrm{CvrFract}}$=$0.6^{+0.1}_{-0.1}$,$\Gamma$=$0.9^{+0.2}_{-0.2}$&\\
&$E_{\mathrm{line}}$=6.7$^{+0.1}_{-0.1}$,${\sigma}_\mathrm{line}$=0.2$^{+0.2}_{-0.1}$&\\
\hline

\end{tabular}
\end{center}
a: $E_{\mathrm{line}}$, ${\sigma}_\mathrm{line}$ and \textit{kT} are in the unit of keV. $N_\mathrm{H}$ is in the unit of $10^{22}\rm{\,cm^{-2}}$. The Galactic foreground extinction is $\rm{6.1\times10^{21}\,cm^{-2}}$ \citep{2016A&A...594A.116H}. \\
}
\label{parameter}
\end{table*}

\subsubsection{Timing Analysis}
We used the {\sc Stingray} \citep{2019ApJ...881...39H} code, which performs $Z^{2}_{n}$ test originally introduced in \citet{1983A&A...128..245B} to detect signals in MOS1, MOS2 and the combined data sets, respectively. We searched in the frequency range 0.0001--1.6\,Hz (MOS sampling time 0.3\,s) with a frequency step size of $1.2\times10^{-6}$\,Hz (1/(10$\times$ExpTime)). Assuming the folded profile can be described by a superposition of three sinusoidal components (n=3), we found the corresponding periods of the most significant signals are around 614.1\,s (7.8$\sigma$) and 1228.3\,s (7.3$\sigma$) in the combined data set (Fig.\,3, \textit{left}). Nevertheless, the most significant signal in MOS2 data is 613.7\,s (1.5$\sigma$). We then folded the X-ray light curve.
According to the MOS1 data set, which has the more significant signal (614.2\,s (5.5$\sigma$)), it seems there are two different peaks within one cycle (Fig.\,3, \textit{right}).

Seven successful \textit{Neutron Star Interior Composition Explorer} \citep[\textit{NICER};][]{2016SPIE.9905E..1HG} observations were made between 2019-09-19 and 2019-11-05. We analysed the data using {\sc HEASoft} 6.28 and CALDB version XTI (20200722). Photons were extracted by \textit{nicerl2} and barycentric corrected by \textit{barycorr}. No significant signals around 614\,s or 1228\,s were detected. Since only about 15\,ks raw data were generated in total and J1745$-$3213 was much fainter than the detection limit of \textit{NICER} \citep{2022AJ....163..130R}, the data are dominated by background noise and the non-detection was not surprising. Based on the longest observation (ID=2200830102, 7ks), the upper limit of the pulsed fraction\citep{1994ApJ...435..362V,2021ApJ...909...33B} is about 19\% with a confidence level of 95\%.

\begin{figure*}
\centering
\includegraphics[width=0.36\textwidth]{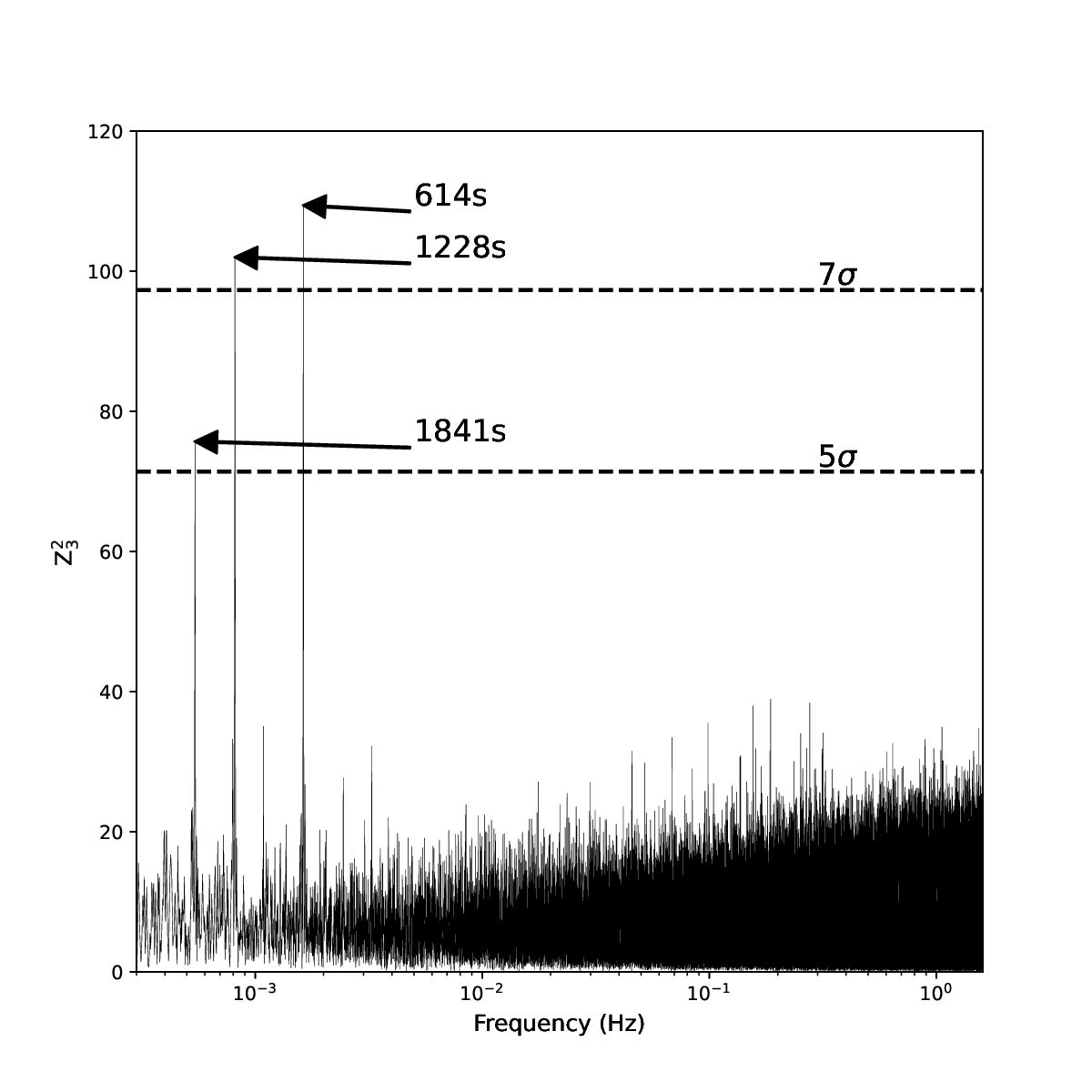}
\includegraphics[width=0.36\textwidth]{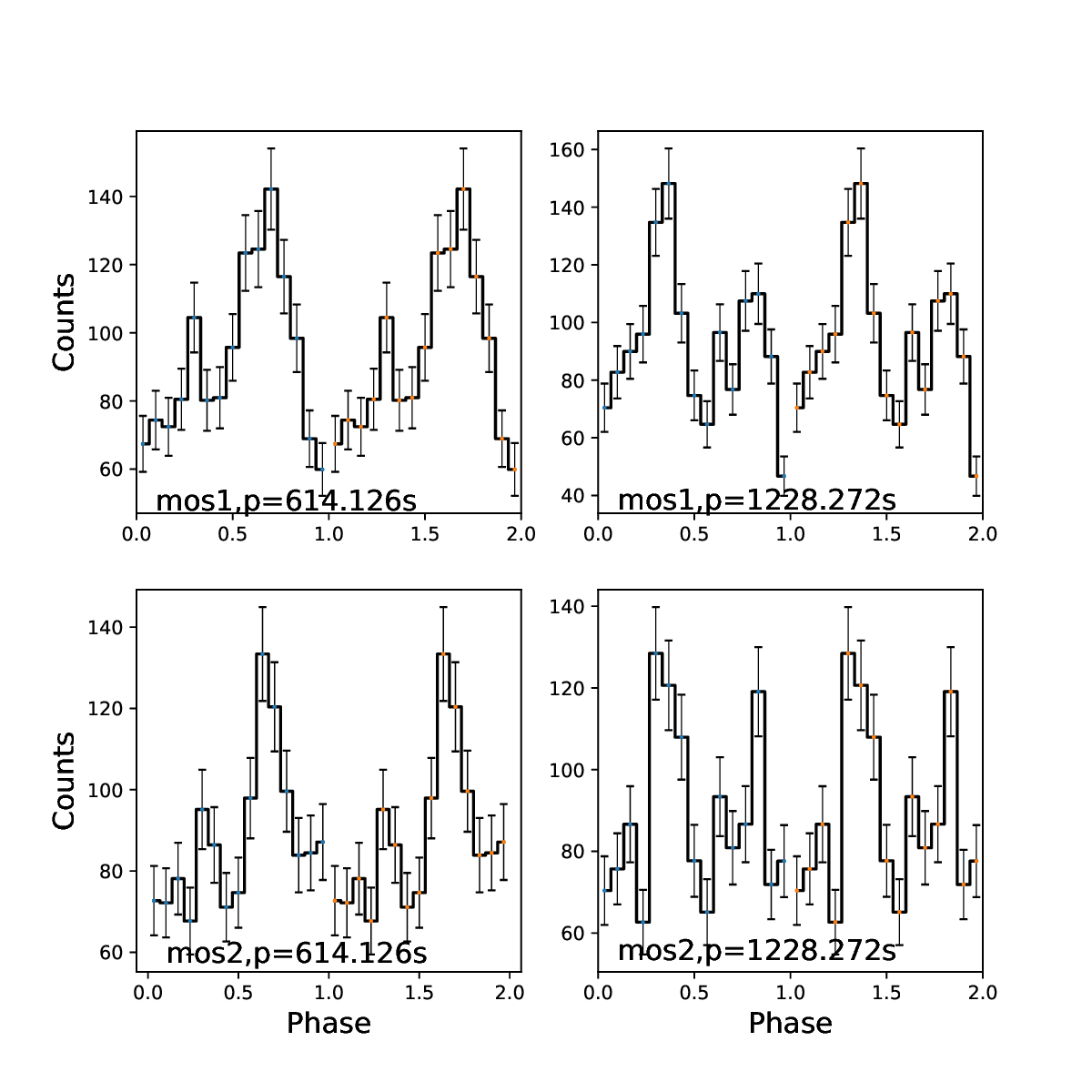}

\caption{Left panel: $Z^{2}_{n}$ statistics obtained from the combined MOS1 and MOS2 data with 3,500 photons. The 614.1\,s (7.8$\sigma$) signal is slightly stronger than the 1228.3\,s signal (7.3$\sigma$) and much stronger than the 1841\,s harmonic signal. Right panels: Comparison of background-subtracted pulse profiles. The MOS1 data set, which has more significant signals, shows two different peaks within 1228\,s. }
\end{figure*}

\subsection{X-ray Observations and Data Analysis}
In order to verify its periodicity and check its spectral shape in the hard X-ray band, we made a quasi-simultaneous
\textit{XMM-Newton} (ObsID=0865510101; PI: Gong) and \textit{NuSTAR} (ObsID=30601021002; PI: Mori) observation on 2021-03-02. This time, the nearby black hole binary H1743$-$322, the target of the old \textit{XMM} observation, was in the quiescent state.

The new \textit{XMM} observation was executed in the full window mode. Its data were reduced by the same procedures in Section 2.1. J1745$-$3213 was on-axis and covered by MOS and PN simultaneously. We adopted a $20''$ source extraction region and obtained 3,600 photons (1,770 PN photons and 1,820 MOS photons, respectively) between 0.3--10\,keV after the GTI correction. Note that the counts had no obvious advantage compared with the old \textit{XMM} observation, but the source extraction regions were different. The background region was simply set to be a homocentric annulus with an inner radius of $20''$ and an outer radius of $40''$ because this time the X source in Fig.\,1 was easily excluded.

The \textit{NuSTAR} data were reduced using the two commands \textit{nupipeline} and \textit{nuproducts} in {\sc HEASoft} 6.28 with the newest calibration files. The source spectra and photons between 3--78\,keV were extracted from r=$40''$ circular regions centered at the centroids of J1745$-$3213$'$s photons, which had $\approx6''$ offsets compared with the \textit{Chandra} position. The spectra were grouped by \textit{grppha} and every spectral bin had a minimum of 25\,counts. The background was selected as a $4'\times2.5'$ rectangular region $\approx 2'$ away. The barycentric correction was applied by setting the input parameter 'barycorr' of \textit{nupipeline} to 'yes'.

\subsubsection{Spectral Analysis}
The MOS and PN spectra were fit simultaneously. If no parameters were fixed, BB yielded the best reduced $\chi^{2}$ and a consistent fit with the previous \textit{XMM} observation (the 1st part of Table.3). Both fits had a BB temperature of $\approx 2$\,keV. If line width was fixed, an \textit{apec} model with a partial covering absorber led to a comparable reduced $\chi^{2}$, with which F-test could not distinguish which model is better. The \textit{XMM} and \textit{NuSTAR} spectra were fit between 0.3--20\,keV simultaneously. As the photons gets harder, the continuum can not be described by BB anymore. Instead, it is characterised by a partial covering \textit{thermal-bremsstrahlung} (Fig.\,2, \textit{middle}) or PL (the 2nd part of Table.3). We selected 3--50\,keV as the hardest band in case of the significant detector background \citep{2016ApJ...826..160H}. The \textit{NuSTAR} spectrum (the 3rd part of Table.3) is described by an absorbed \textit{thermal-bremsstrahlung} (Fig.\,2, \textit{right}) with an unabsorbed flux and luminosity of $3.8^{+0.4}_{-0.4}\times\rm{10^{-12}\,erg\,cm^{-2}\,s^{-1}}$ and $4.6^{+0.5}_{-0.5}\times10^{32}\times$(D/1 kpc)$^{2}$ erg s$^{-1}$, respectively. A slightly inferior model is absorbed \textit{apec}. Adding a partial covering component to either of the two models would yield a worse reduced $\chi^{2}$ and let the partial covering factor hit its maximum boundary 1.0.

\begin{table*}
\scriptsize{
\caption{X-Ray Spectral Fits Based on New \textit{XMM} and \textit{NuSTAR} Data}
\begin{center}
\begin{tabular}{lll}
\hline\hline
Model & Parameters&red.$\chi^{2}$(dof) \\
\hline
\hline
&MOS1+MOS2+PN\,(0.3-10\,keV)&\\
\hline
xstbabs*xspcfabs*(xsbremss+xsgaussian)&$N_\mathrm{H}1$=$2.8^{+0.3}_{-0.3}$,$N_\mathrm{H}2$=$13.8^{+4.6}_{-4.6}$&1.10(147)\\
&$f_{\mathrm{CvrFract}}$=$0.6^{+0.08}_{-0.08}$,\textit{kT}=$12.7^{+4.7}_{-4.7}$&\\
&$E_{\mathrm{line}}$=6.7$^{+0.05}_{-0.05}$,${\sigma}_\mathrm{line}$=0.2$^{+0.05}_{-0.05}$&\\
\hline
xstbabs*xsapec&$N_\mathrm{H}$=$4.2^{+0.2}_{-0.2}$,\textit{kT}=$14.1^{+1.8}_{-1.8}$&1.57(152)\\
\hline
xstbabs*xspcfabs*(xsapec+xsgaussian)&$N_\mathrm{H}1$=$2.7^{+0.3}_{-0.3}$,$N_\mathrm{H}2$=$10.9^{+3.1}_{-3.1}$&1.08(148)\\
&$f_{\mathrm{CvrFract}}$=$0.6^{+0.07}_{-0.07}$,\textit{kT}=$14.3^{+2.0}_{-2.0}$&\\
&$E_{\mathrm{line}}$=6.5$^{+0.04}_{-0.04}$,${\sigma}_\mathrm{line}$=0.05(fixed)\\
\hline
xstbabs*xspcfabs*(powlaw1d+xsgaussian)&$N_\mathrm{H}1$=$3.1^{+0.3}_{-0.3}$,$N_\mathrm{H}2$=$13.1^{+3.7}_{-3.7}$&1.10(147)\\
&$f_{\mathrm{CvrFract}}$=$0.6^{+0.08}_{-0.08}$,$\Gamma$=$1.9^{+0.2}_{-0.2}$&\\
&$E_{\mathrm{line}}$=6.7$^{+0.05}_{-0.05}$,${\sigma}_\mathrm{line}$=0.2$^{+0.05}_{-0.05}$&\\
\hline
xstbabs*(xsbbody+xsgaussian)&$N_\mathrm{H}$=$1.3^{+0.1}_{-0.1}$,\textit{kT}=$1.9^{+0.06}_{-0.06}$&1.09(149)\\
&$E_{\mathrm{line}}$=6.7$^{+0.04}_{-0.04}$,${\sigma}_\mathrm{line}$=0.2$^{+0.05}_{-0.05}$&\\
\hline
\hline
&\textit{XMM}+\textit{NuSTAR}\,(0.3-20\,keV)&\\
\hline
xstbabs*xsbremss&$N_\mathrm{H}$=$3.9^{+0.2}_{-0.2}$,\textit{kT}=$33.0^{+5.4}_{-5.4}$&1.46(253)\\
\hline
xstbabs*(xsbremss++xsgaussian)&$N_\mathrm{H}$=$3.5^{+0.2}_{-0.2}$,\textit{kT}=$38.8^{+7.3}_{-7.3}$&1.09(250)\\
&$E_{\mathrm{line}}$=6.6$^{+0.04}_{-0.04}$,${\sigma}_\mathrm{line}$=0.3$^{+0.05}_{-0.05}$&\\
\hline
xstbabs*xspcfabs*(xsbremss+xsgaussian)&$N_\mathrm{H}1$=$2.6^{+0.3}_{-0.3}$,$N_\mathrm{H}2$=$9.2^{+3.6}_{-3.6}$&1.01(248)\\
&$f_{\mathrm{CvrFract}}$=$0.5^{+0.1}_{-0.1}$,\textit{kT}=$21.8^{+4.0}_{-4.0}$&\\
&$E_{\mathrm{line}}$=6.6$^{+0.04}_{-0.04}$,${\sigma}_\mathrm{line}$=0.2$^{+0.05}_{-0.05}$&\\
\hline
xstbabs*xsapec&$N_\mathrm{H}$=$3.9^{+0.2}_{-0.2}$,\textit{kT}=$24.7^{+2.8}_{-2.8}$&1.34(253)\\
\hline
xstbabs*(xsapec+xsgaussian)&$N_\mathrm{H}$=$3.5^{+0.2}_{-0.2}$,\textit{kT}=$37.3^{+7.0}_{-7.0}$&1.10(250)\\
&$E_{\mathrm{line}}$=6.6$^{+0.04}_{-0.04}$,${\sigma}_\mathrm{line}$=0.3$^{+0.05}_{-0.05}$&\\
\hline
xstbabs*xspcfabs*(xsapec+xsgaussian)&$N_\mathrm{H}1$=$2.4^{+0.4}_{-0.4}$,$N_\mathrm{H}2$=$6.7^{+2.6}_{-2.6}$&1.03(248)\\
&$f_{\mathrm{CvrFract}}$=$0.5^{+0.1}_{-0.1}$,\textit{kT}=$25.4^{+4.2}_{-4.2}$&\\
&$E_{\mathrm{line}}$=6.5$^{+0.05}_{-0.05}$,${\sigma}_\mathrm{line}$=0.2$^{+0.06}_{-0.06}$&\\
\hline
xstbabs*xspcfabs*(powlaw1d+xsgaussian)&$N_\mathrm{H}1$=$3.1^{+0.3}_{-0.3}$,$N_\mathrm{H}2$=$12.4^{+3.1}_{-3.1}$&1.01(248)\\
&$f_{\mathrm{CvrFract}}$=$0.6^{+0.05}_{-0.05}$,$\Gamma$=$1.9^{+0.1}_{-0.1}$&\\
&$E_{\mathrm{line}}$=6.7$^{+0.04}_{-0.04}$,${\sigma}_\mathrm{line}$=0.2$^{+0.05}_{-0.05}$&\\
\hline
xstbabs*(xsbbody+xsgaussian)&$N_\mathrm{H}$=$0.9^{+0.1}_{-0.1}$,\textit{kT}=$2.2^{+0.04}_{-0.04}$&1.52(250)\\
&$E_{\mathrm{line}}$=6.6$^{+0.05}_{-0.05}$,${\sigma}_\mathrm{line}$=0.2$^{+0.05}_{-0.05}$&\\
\hline
\hline
&FPMA+FPMB\,(3-50\,keV)&\\
\hline
xstbabs*xsbremss&$N_\mathrm{H}$=$5.2^{+1.4}_{-1.4}$,\textit{kT}=$23.6^{+3.8}_{-3.8}$&1.02(109)\\
\hline
xstbabs*(xsbremss++xsgaussian)&$N_\mathrm{H}$=$2.8^{+1.5}_{-1.5}$,\textit{kT}=$35.7^{+8.7}_{-8.7}$&0.85(106)\\
&$E_{\mathrm{line}}$=6.5$^{+0.1}_{-0.1}$,${\sigma}_\mathrm{line}$=0.4$^{+0.1}_{-0.1}$&\\
\hline
xstbabs*xsapec&$N_\mathrm{H}$=$4.1^{+1.2}_{-1.2}$,\textit{kT}=$26.2^{+3.6}_{-3.6}$&1.00(109)\\
\hline
xstbabs*(xsapec+xsgaussian)&$N_\mathrm{H}$=$2.2^{+1.4}_{-1.4}$,\textit{kT}=$38.1^{+9.0}_{-9.0}$&0.88(106)\\
&$E_{\mathrm{line}}$=6.4$^{+0.2}_{-0.2}$,${\sigma}_\mathrm{line}$=0.4$^{+0.2}_{-0.2}$&\\
\hline
xstbabs*ipm&$N_\mathrm{H}$=$6.4^{+3.1}_{-3.1}$,$f_{\mathrm{fall\_height}}$=$7.0^{+2.8}_{-2.8}$,$f_{\mathrm{m\_wd}}$=$1.0^{+0.2}_{-0.2}$&1.01(108)\\
\hline

\end{tabular}
\end{center}
}
\label{parameter}
\end{table*}

\subsubsection{Timing Analysis}
The MOS and PN sampling times of the new observation were 2.6\,s and 73.4\,ms, respectively. For the PN data set, we searched in the frequency range 0.0001--6.8\,Hz with a frequency step size of $2.0\times10^{-6}$\,Hz (1/(10$\times$ExpTime)) via $Z^{2}_{n}$ test (n=3). The two signals were confirmed (Fig.\,4, \textit{left}). Nevertheless, we found the corresponding period of the most significant signal is around 613.5\,s (4.7$\sigma$), which is weakened by strong red noise. No shorter significant signals were discovered. For the combined MOS data set, we searched in the frequency range 0.0001--0.19\,Hz and found the corresponding period of the most significant signal is around 613.8\,s (7.0$\sigma$). The 1,228\,s signal was both weaker in the two cases (Fig.\,4, \textit{left}).
Considering in this observation J1745$-$3213 was on-axis and had a small source extraction region, and especially the nearly identical peaks we obtained in the MOS data (Fig.\,4, \textit{right}), we take the $613.8^{+1.0}_{-0.8}$\,s \citep{2019ApJ...881...39H} signal as the fundamental period tentatively.

Surprisingly, no signals were detected in the {NuSTAR} photons no matter a r=$40''$ or a r=$20''$ extraction region was adopted. The non-detection was not the result of the choice of energy band because 3--10\,keV and 3--78\,keV were both tried. We attribute the non-detection to the high background of J1745$-$3213 in the {NuSTAR} observation.

\begin{figure*}
\centering
\includegraphics[width=0.4\textwidth]{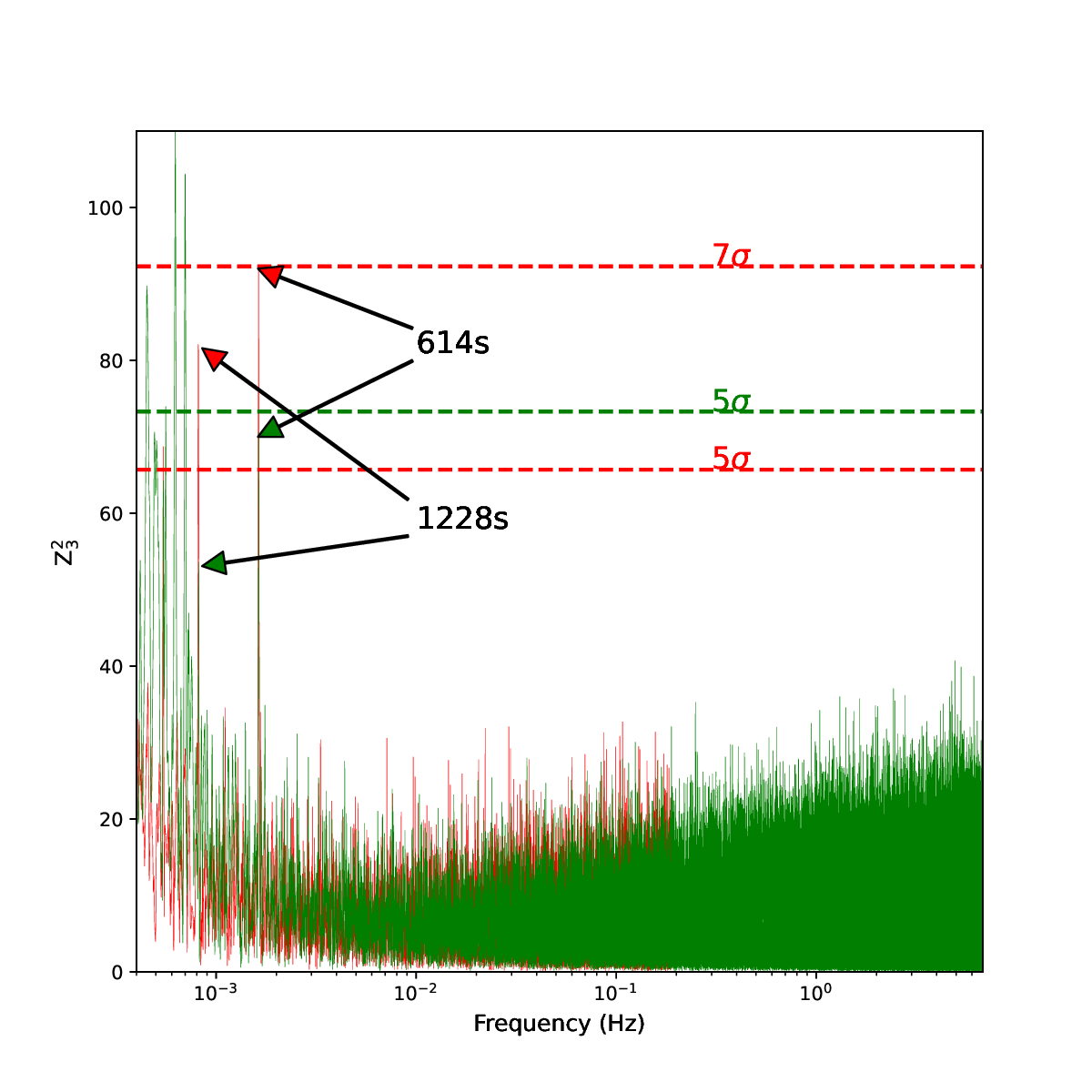}
\includegraphics[width=0.4\textwidth]{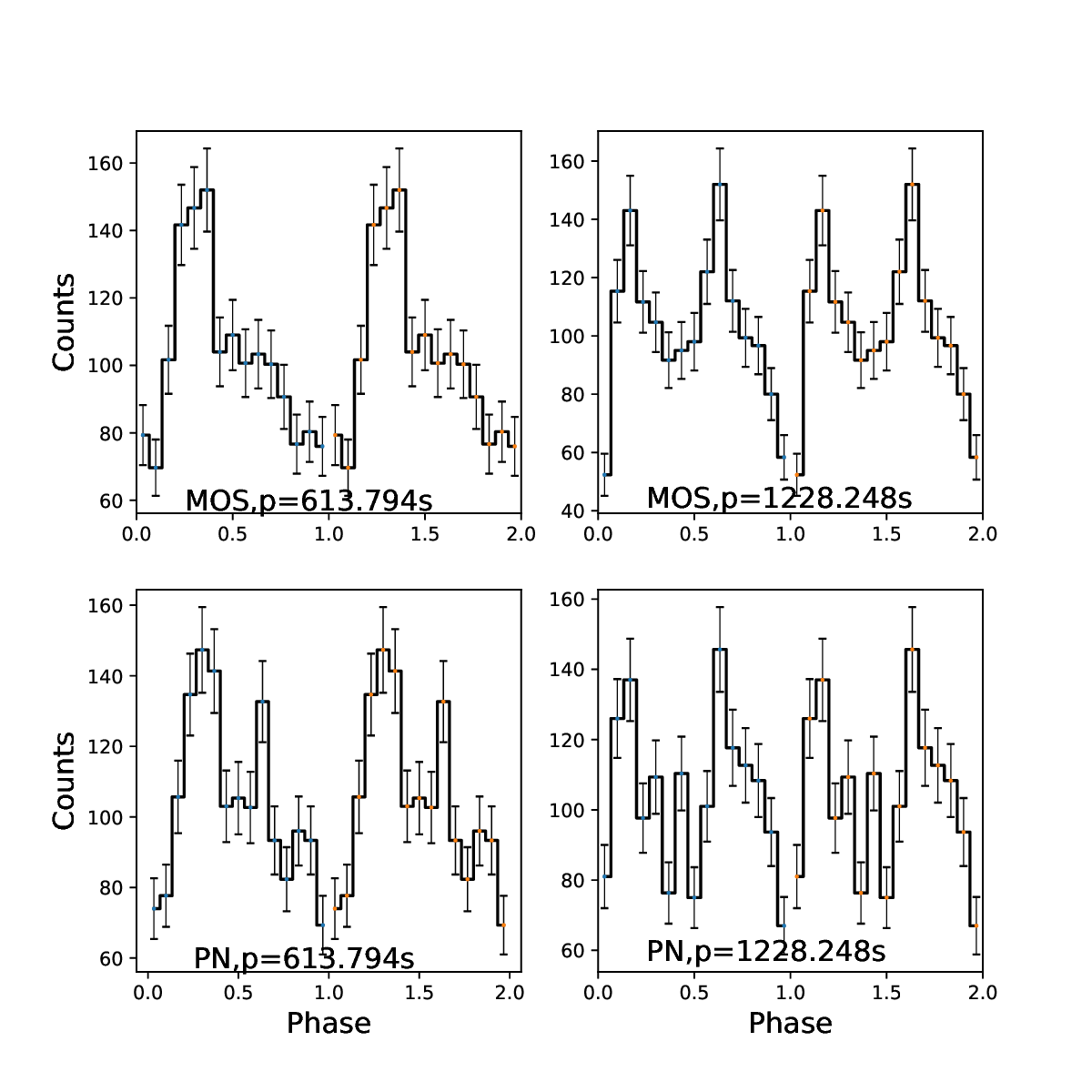}

\caption{Left panel: $Z^{2}_{n}$ statistics obtained from the PN (green) and combined MOS (red) data sets. The 613.8\,s (7.0$\sigma$) signal is slightly stronger than the 1228.2\,s signal (6.3$\sigma$) for MOS. Right panels: Comparison of background-subtracted pulse profiles. For MOS, which does not have strong red noise, the difference between two peaks within 1228\,s is insignificant, favouring 614\,s as the fundamental period. The pulsed fraction (0.3--10\,keV) is roughly $\frac{150-110}{150+110}$=15.4\%.}
\end{figure*}

\section{Archival Optical Data Analysis}
J1745$-$3213 is fortuitously located in a field with archival VLT observations. Its VLT R-band image (Fig.\,1, \textit{right}) displays a patchy environment and a nonuniform extinction near J1745-3213. We aligned the astrometry of the VLT-VIMOS image onto the \textit{Chandra} image using four \textit{Tycho} stars in the field of view. No optical point sources brighter than the limit of $\rm{R}=25\,$mag\footnote{\url{https://www.eso.org/observing/etc/bin/gen/form?INS.NAME=VIMOS+INS.MODE=imaging}} were detectable within $1''$ of the X-ray position.

However, the extinction towards the Galactic center is very high. Assuming $\rm{R_V}=3.1$ and adopting a moderate $N_\mathrm{H}$ (2.0$\times 10^{22}\rm{cm^{-2}}$) based on Table.2 and Table.3, we obtained $\rm{A(V)}=11.2\,mag$ and $\rm{A(R)}\approx9.6\,mag$ ($\lambda_{eff}$=6,312\,\AA)\footnote{\url{http://www.dougwelch.org/Acurve.html}} using the extinction model from \cite{1989ApJ...345..245C}. Therefore, we can exclude a G5-type giant star within 11\,kpc, an M5-type giant star within 37\,kpc, an A0-type main sequence star within 9\,kpc, an A5-type main sequence star within 5.3\,kpc, and a B8-type main sequence star within 13.4\,kpc (Table 15.7, Allen$¡¯$s Astrophysical Quantities, 4th Ed.). But, if we adopted a smaller $N_\mathrm{H}$ (1.0$\times 10^{22}\rm{cm^{-2}}$), an M0-type main sequence star within 3.6\,kpc, a G2-type main sequence star within 16.8\,kpc and an A5-type main sequence star within 50\,kpc can be excluded.

We note there is a reported cold clump PGCC\,G357.16$-$1.66 \citep{2016A&A...594A..28P} in the same direction. With semi-major and semi-minor axes of $4.5'\times3.4'$, and an angular separation of just $40''$, the extinction may be enhanced by the clump.
\section{Radio Observations and Data Analysis}
\subsection{ATCA}
Since the optical data showed no source at the location of the Chandra position, we requested and received radio observations on the ATCA.
J1745$-$3213 was observed over a period of four epochs: 2018-12-22, 2018-12-23, 2018-12-27 and 2019-01-03 for a total integration of $\simeq$11.5 hours on source. For each observing epoch, the array was in the 1.5D configuration with a minimum baseline length of 107\,m and a maximum baseline length of nearly 4,438\,m. We utilise the Compact Array Broadband Backend \citep[CABB,][]{2011MNRAS.416..832W} with our observations being taken at a centre frequency of 2,100\,MHz with 2\,GHz of instantaneous bandwidth. Continuum data was recorded in 2,048 independent channels of 1\,MHz bandwidth each. The flux and bandpass calibrator used for these observations was PKS\,1934$-$638 while PKS\,1759$-$39 was used as a phase calibrator.

Data were reduced, calibrated and imaged using the {\sc miriad} software package \citep{1995ASPC...77..433S} with standard routines. Flagging of the data was largely done with the automated task \textit{pgflag}.
We made naturally-weighted total intensity (Stokes $I$) mutifrequency images of 900\,MHz of bandwidth, centred on 1,730 and 2,550\,MHz (Fig.\,5). The resultant images have a measured root-mean-squared noise ($\sigma$) of 24 and $20\,\mu$Jy\,beam$^{-1}$ for the 1,730\,MHz and 2,550\,MHz images, respectively.

Interestingly, the source was detected in the upper band, and not the lower band (Fig.\,5). The upper-band image centred at 2,550\,MHz shows a source at the centre of the imaged field with a peak intensity of $128\,\mu$Jy\,beam$^{-1}$, equivalent to $6.4\sigma$, and suggests a strong positive spectral index of $\alpha > 0.51$ where $S_{\nu} \propto \nu^{\alpha}$. Imaging the entire 2\,GHz of bandwidth results in a detection of lesser significance ($5.1\sigma$) when compared to the upper-band image.

\begin{figure*}
\centering
\includegraphics[width=0.4\textwidth]{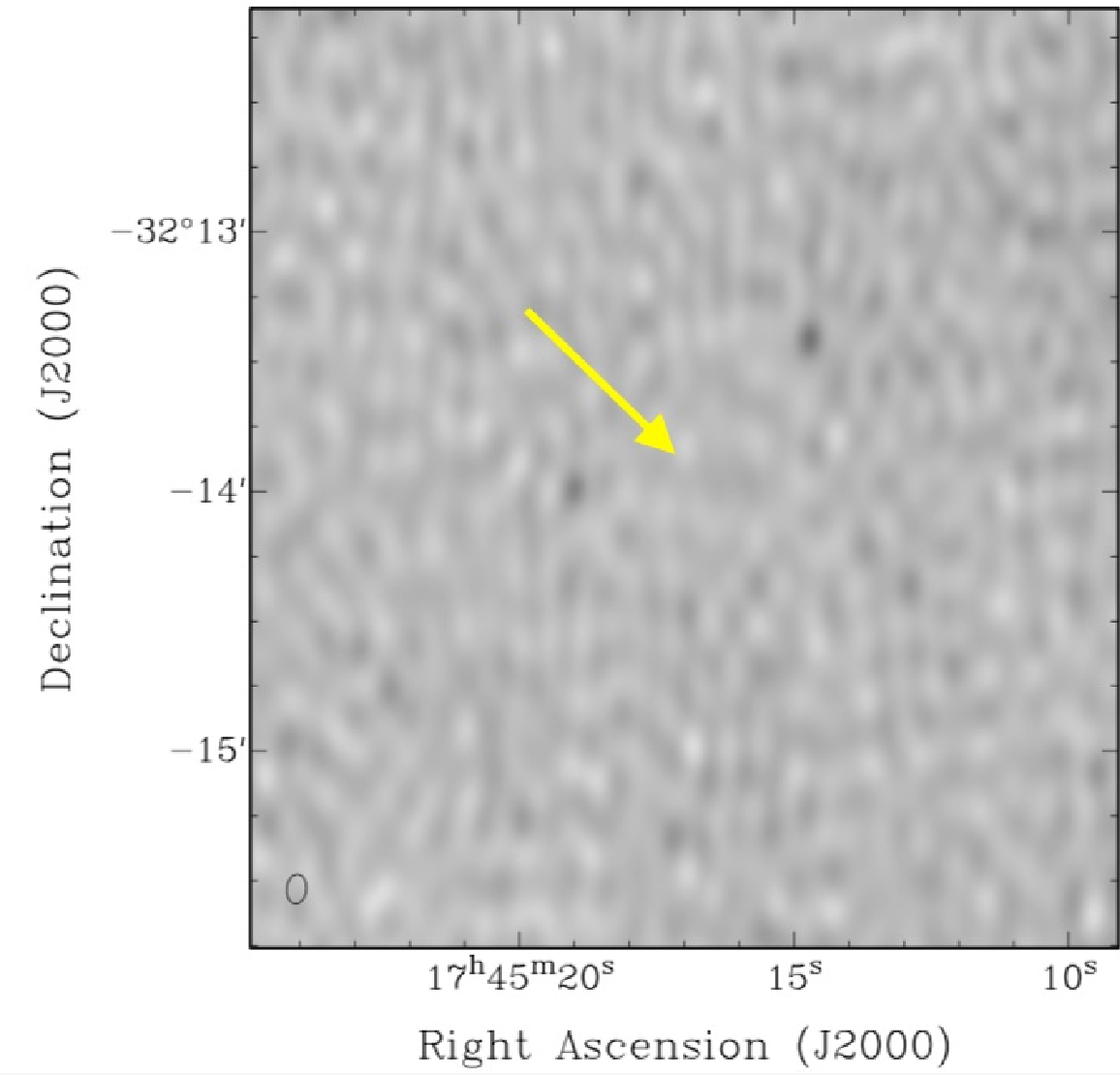}
\includegraphics[width=0.4\textwidth]{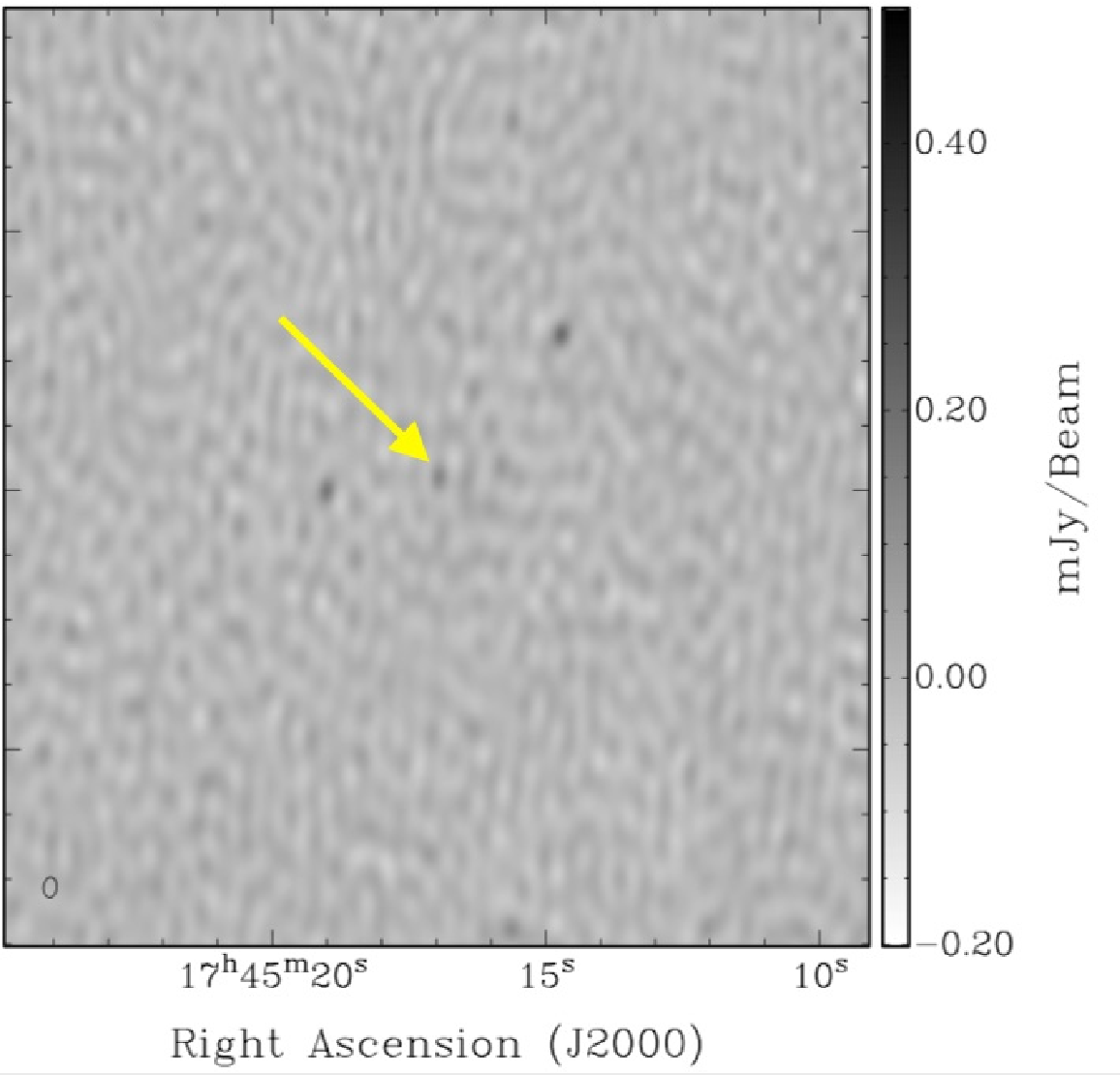}
\caption{Total intensity (Stokes $I$) radio continuum images towards J1745 at 1,730\,MHz (\textit{left}) and 2,550\,MHz (\textit{right}). The RMS noise of the images is $24$ and $20\,\mu$Jy\,beam$^{-1}$, respectively. The two images have the same flux scale and are gridded to the same map. The synthesised beam is shown in the lower left corner.}
\label{fig:atca_stokesI}
\end{figure*}

The existence of a signal in only the higher-frequency band is unlikely due to a difference in sensitivity between the two bands, because the number of flagged visibilities due to radio frequency interference (RFI) between the two bands results in the higher-frequency band being only about 10\% more sensitive than the lower-frequency band. To test whether the radio emission at the source position was an artefact related to either RFI or some other effect, we systematically removed antennae and observational epochs. In all cases the radio source remained, although we note that removing single epochs provided different source intensity, suggesting J1745$-$3213 is mildly radio variable. Despite radio emission remaining for all tests, we do remain somewhat cautious of the source detection.

Out of caution, we followed up our initial set of radio observations with a further ATCA radio observation taken at central frequencies of 5.5 and 9\,GHz. This later radio observation was recorded simultaneously with a bandwidth of 2\,GHz at each frequency. It was made on 2020-06-18 when the array was in the 1.5C configuration. Calibration, data reduction and analysis followed the same steps as reported above.

J1745$-$3213 was not detected in either the 5.5 or 9\,GHz band. We measured an RMS noise of 17 and $15\,\mu$Jy\,beam$^{-1}$ for 5.5\,GHz and 9\,GHz, respectively, over the source position. While tests suggest that the detected radio emission from our 2018/2019 data was likely real, the non-detection in the lower-half of the band, and the non-detection at later times (at a higher frequency) means that the detection at 2,550\,MHz should be regarded as preliminary.

\subsection{VLA}
We also performed follow-up radio observations using the VLA. J1745$-$3213 was observed for six hours on four separate dates (Project Code=21A-156; PI: Gong). The first two observations measured the spectrum of J1745$-$3213 across five radio bands from 2 to 26\,GHz during two separate one hour
scheduling blocks. One for the three lowest frequencies and the other
for the two remaining high frequencies. The exposures in each band
range between 705\,s to 948\,s duration. The third and fourth
observations were two hours each in the C and X bands (4--8 and 8--12\,GHz, respectively) with an on-source time of about 1.6 hours each. During
these observations, the VLA cycled between the target J1745$-$3213 and
the phase calibrator J1744$-$3116 about every 570\,s with about 525\,s
being on target. Each observation used the 3-bit samplers for wideband
coverage. The radio sources J1331$+$305 (=3C286) and J1744$-$3116 were
used as the flux and polarization, and the phase and gain calibrators
for each observation, respectively. No polarization leakage
calibrators were observed, because the cross-polarization is $<$1\% and
varies slowly over several months. This accuracy
is sufficient for our needs. Table.4 gives a log of the observations
and the measured flux densities in each band. No radio source was
detected at the location of J1745$-$3213 in any of the observations (e.g., Fig.\,6).

The data were calibrated using version 5.6.3 of the  {\sc CASA} (Common Astronomy Applications Software) calibration pipeline \citep{2007ASPC..376..127M}. The imaging application \textit{tclean} is used to generate IQUV and RRLL images of each target scan to check for any source confusion and
radio frequency interference. None were found. The flux densities are
measured by fitting a point source to the UV data using the Julia
programming language package \textit{Visfit} \citep{2022AJ....163...58B}. The package uses a
box-constrained Levenberg-Marquardt algorithm to minimize the model
residuals. The position of the point source is constrained to be
within a $4''\times4''$ region centered on the X-ray source position.

\begin{table*}
\begin{center}
\caption{VLA Observation List}
\begin{tabular}{llllll}
\hline\hline
Date   &   Band  & Freq  &  $\Delta$RA$^{a}$     &    $\Delta$Dec$^{a}$  &    I\\
&&(GHz)&(arcsecond)&(arcsecond)&($\mu$Jy\,beam$^{-1}$)\\
2021-06-04T06:25:25Z&  S&    3.0&  -2.00$\pm4.00$&   2.00$\pm9.00$ &   14$\pm42$  \\
2021-06-04T07:10:45Z&  C&    6.0&   0.89$\pm0.05$&  -0.61$\pm0.11$ &  -50$\pm33$  \\
2021-06-04T06:03:00Z&  X&    10.0&   0.39$\pm0.19$ &   1.44$\pm2.82$ &   10$\pm57$  \\
2021-06-12T05:28:3Z5&  Ku&   15.0&   1.59$\pm0.48$ &  -0.93$\pm0.87$ &   29$\pm36$  \\
2021-06-12T05:09:25Z&  K&    22.0&  0.53$\pm0.69$&  -1.32$\pm1.15$&  -27$\pm63$  \\
\hline

\end{tabular}
\end{center}
a: The distance in arcseconds between the fitted point source position and the known X-ray position of J1745$-$3213. The search for a point source is constrained to a $4''\times4''$ region. \\
\end{table*}

\begin{figure*}
\centering
\includegraphics[width=0.4\textwidth]{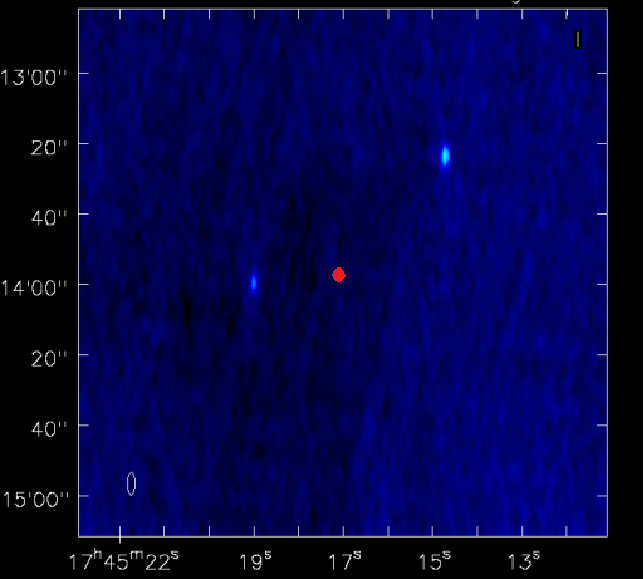}
\caption{J1745$-$3213 is supposed to be located at the center (red spot) of this C-band image, which has only two irrelevant sources. The RMS error for the image is about 5\,$\mu$Jy\,beam$^{-1}$.}

\end{figure*}

\section{The Nature of J1745$-$3213}
Based on the 614\,s signal, the 6.7-keV iron feature, non-detection of an optical counterpart and its small Galactic latitude ($l,\ b=357.14866^\circ,\ -1.65597^\circ$), it is safe to conclude that J1745$-$3213 is a Galactic compact object. Nevertheless, as we point out in the first section, it is not necessarily easy to distinguish between a rotational period and an orbital period. Being devoid of an optical counterpart makes a secure identification difficult. Below, we mainly discuss its possibility of being  an X-ray binary or a white dwarf binary.

The signal may be interpreted as the rotational period of an accreting pulsar. Pulsars in LMXBs typically spin faster ($\rm{P_{\mathrm{spin}}}<200\,$s) than J1745$-$3213 due to high accretion torques. Even the exceptions to the LMXB fast spin rule, symbiotic LMXBs \citep[$\rm{P_{\mathrm{spin}}}>100\,$s, e.g.,][]{2007A&A...470..331M} are unlikely too due to their rarity and the presence of M-type giants as companions. The parameter space of a NS-HMXB scenario is quite restricted due to the optical and X-ray faintness of J1745$-$3213. Being a HMXB would require an $L_{\mathrm{X}}$ much higher than $1.9\times10^{32}\times$(8/1 kpc)$^{2}$ erg s$^{-1}$ \citep{2002A&A...391..923G}.

The signal may be the orbital period of a LMXB, or further, an ultra-compact X-ray binary (UCXB), which belongs to LMXBs with extremely short orbital periods \citep[$<$1 hour, e.g.,][]{2007A&A...465..953I}. Due to their short orbital periods, which lead to small Roche lobes, the donors of UCXBs are supposed to be hydrogen-stripped stars or white dwarfs. They tend to be located in globular clusters and the Galactic bulge, while about 40\% \citep{2013ApJ...768..184H} are just Galactic field UCXBs. A small donor would make J1745$-$3213 optically faint \citep{2007A&A...469.1063I, 2009A&A...506..857I}. The existence of an iron feature suggests a He-rich donor \citep{2014MNRAS.442.2817K}. One may argue that a persistently and extremely low Eddington ratio ($\approx$0.01$\%$ even if $\rm{D}$=10\,kpc and M=1.4\,$\rm M_{\odot}$) for J1745$-$3213 is incompatible with a supposedly high $L_{\mathrm{X}}$ driven by gravitational radiation \citep{2004ApJ...607L.119B}. But, it may be a transient UCXB \citep{2013ApJ...768..184H} and a very faint X-ray binary \citep{2015MNRAS.447.3034H}, which both have high concentrations near the Galactic center \citep{2021MNRAS.501.2790B}. If the BB choice in the \textit{XMM} band is correct, then J1745$-$3213 is a NS-UCXB because the X-ray spectrum would be much harder for a black hole X-ray binary with such a low $L_{\mathrm{X}}$. The low $L_{\mathrm{X}}$ can be explained by a small accretion flow to a neutron star limited by the propeller effect and a small disk \citep{2015MNRAS.447.3034H,2018MNRAS.475.2027V}. The BB emission may originate from a hot spot with a radius about 11.3$\times$(D/1\,kpc)\,m \citep{1973ApJ...184..271L}, or more likely, from two antipodal spots on the surface of J1745$-$3213. Similar spots appeared on the surfaces of the NS X-ray binaries 1E1743.1$-$2843 \citep[\textit{kT}=1.8\,keV,][]{2016ApJ...822...57L} and Cir X-1 \citep[\textit{kT}=1.6\,keV,][]{2020ApJ...891..150S}. The hard band X-ray spectrum of J1745$-$3213 is consistent with an IP scenario, but not exclusive to it \citep[e.g.,][]{2014ApJ...797...92C}. If the fundamental period is 1228\,s, one piece of evidence against the UCXB scenario is its possible pulse profile. A 180\,$^{\circ}$-separated double-peaked pulse profile usually corresponds to a rotational signal from compact objects like IPs and accreting pulsars \citep{2010A&A...520A..76A}. A standard example of sources with this sort of pulse profile is black widow pulsars, but a 180\,$^{\circ}$-separated double-peaked pulse profile would require a fine-tuning of the geometries of intra-binary shocks \citep{2016ApJ...828....7R}. Although orbital modulations of X-ray binaries rarely generate this kind of profile, positive examples do exist, e.g., the HMXB Vela X-1 \citep{2016A&A...588A.100M}. If the radio observations in the 1--3\,GHz range are accurate, the emission contrast between the lower and higher band of 1--3\,GHz could be explained by self-absorbed jets \citep{2006MNRAS.367.1083K} naturally.

The spin of polars is usually synchronized or slightly asynchronous to their orbital motion. Except AM CVn systems, most of the CVs have orbital periods longer than 80\,mins \citep{2017PASP..129f2001M}. For the white dwarf binary scenario, considering the length of the 614\,s signal, we interpret it as the rotational period of an IP or the orbital period of an AM CVn system. IPs \citep{2003cvs..book.....W,2017PASP..129f2001M} are a subclass of CVs with moderate magnetic fields strong enough to truncate accretion discs, but weaker than the magnetic fields of polars, which prevent the formation of discs. The accretion stream would follow magnetic filed lines and form a shock when it collides with the magnetic poles of white dwarfs. IPs can show Fe K$_{\alpha}$ complex \citep{2004MNRAS.352.1037H}, but this is almost ubiquitous in X-ray sources. A secure identification requires a spectroscopic observation in the optical band \citep[e.g.,][]{2021ApJ...914...85H}, which is impossible for J1745$-$3213. Arguments in favor of an IP identity include: 1) the hard band X-ray spectral shape is consistent with IPs \citep{2017PASP..129f2001M}. Sources with thermal bremsstrahlung temperatures in the range of dozens of keV are often identified to be IPs \citep{2016MNRAS.460..513T,2016MNRAS.461..304C,2021ApJ...914...85H,2022MNRAS.511.4582C},
2) at least eight IPs show 180\,$^{\circ}$-separated double-peaked pulse profiles \citep{1999A&A...347..203N,2002A&A...384..195N,2012MNRAS.420.3350N} generated by two-pole accretion, which is good news for an IP scenario if 1228\,s is fundamental, and 3) IPs have late-type companions. The strongest counter-argument to the IP nature is the tentative radio detection by ATCA. IPs are not strong radio emitters \citep{2017AJ....154..252B, 2020AdSpR..66.1226B}. J1745$-$3213 even showed evidence of radio emission with a positive spectral index. But follow-up radio observations could not confirm it. Therefore, either that detection is an artefact or J1745$-$3213 is transient in the radio band. If J1745$-$3213 is indeed an IP, using the PSR model derived by \citet{2019MNRAS.482.3622S}, we can estimate a $1.0^{+0.2}_{-0.2}\,\rm M_{\odot}$ mass (last line of Table.3) for J1745$-$3213, which is compatible with the typical mass of an IP \citep{2020MNRAS.498.3457S}. For AM CVn systems in particular, they can have Fe K$_{\alpha}$ emission \citep{2005A&A...440..675R} and usually can be modelled by optically thin thermal plasma emission \citep{2007ASPC..372..425R}. Nevertheless, for those short-period AM CVn systems, they are modelled by BB emission with temperatures of dozens of eV \citep{2004MNRAS.349..181N,2010PASP..122.1133S}.

In summary, properties like the hard thermal bremsstrahlung spectrum, double-peaked X-ray pulse profile (if 1228\,s is fundamental), faint X-ray luminosity, and an invisible companion in the optical band are naturally compatible with an IP origin. Considering either of these properties can be cracked, the nonexistence of an optical counterpart, and the possibility that the J1745$-$3213 is transient in the radio band, based on the current data, we argue the second choice for J1745$-$3213 is an ultra-compact X-ray binary. A deep search for an infrared counterpart may help reveal its nature.

\section*{Acknowledgements}
GH would like to thank J.\ F. Kaczmarek and Paul Barrett for radio observations, data analysis and writing the radio sections.

GH would also like to thank Matteo Bachetti, Roberto Soria, Sean Lake, Kaya Mori, Frank Haberl, George Hobbs and Charles Hailey for proposing observation time, data analysis and revising the paper. Thomas Russell, Liu Zhu, Ge Mingyu, Stefania Carpano, Kim Page, Mason Ng, Keith Arnaud, Feng Ye, Yve Schutt, Cui Kaiming and Jin Chichuan helped analysing the data directly or indirectly. GH is grateful to Victor Doroshenko, Valery Suleimanov and Aarran Shaw for their guidance about intermediate polar.

GH would also like to thank the anonymous reviewers for their careful reading of the paper and presenting constructive suggestions.

The Australia Telescope is funded by the Commonwealth of Australia for operation as a National Facility managed by CSIRO. The National Radio Astronomy Observatory is a facility of the National Science Foundation operated under cooperative agreement by Associated Universities, Inc.

This research has made use of software provided by the \textit{Chandra} X-ray Center (CXC) in the application packages {\sc CIAO} and {\sc Sherpa}.
This research is based on observations obtained with \textit{XMM-Newton}, an ESA science mission with instruments and contributions directly funded by ESA Member States and NASA. This research has made use of the \textit{NuSTAR} Data Analysis Software (NuSTARDAS) jointly developed by the ASI Space Science Data Center (SSDC, Italy) and the California Institute of Technology (Caltech, USA).

This work is supported by the National Natural Science Foundation of China (Grants no. 11703041).

\software{CIAO \citep[v4.14;][]{2006SPIE.6270E..1VF}, HEASoft (v6.28), CASA \citep[v5.6.3;][]{2007ASPC..376..127M}, Stingray \citep{2019ApJ...881...39H}, SAS \citep[v19;][]{2004ASPC..314..759G}, Miriad \citep{1995ASPC...77..433S}}

\end{document}